\documentclass[10pt,conference]{IEEEtran}
\IEEEoverridecommandlockouts

\renewcommand{\baselinestretch}{0.96}

\usepackage{algorithm} % The algorithm environment
\usepackage{algorithmicx} % The good pseudocode package
\usepackage[noend]{algpseudocode} 
\usepackage{mathtools}  % Contains mathclap, which removes the effects of long 
\newcommand{\para}[1]{\smallskip \textit{#1.}}

\usepackage{graphicx}

\usepackage{amsmath,amsfonts}
\newcommand{\isdef}{:=}

\usepackage{pdfrender}

\usepackage{pifont}

\usepackage{array,multirow,graphicx}
\usepackage{float}

\newtheorem{theorem}{Theorem}[section]
\newtheorem{problem}[theorem]{Problem}

\newtheorem{property}[theorem]{Property}

\newcommand{\figref}[1]{Fig.~\ref{#1}}
\newcommand{\multifigref}[2]{Fig.~\ref{#1} (#2)}

\newcommand{\agent}[1]{a_{#1}}
\newcommand{\agents}[0]{\boldsymbol{A}}
\newcommand{\noagents}[0]{c}

\newcommand{\floorplanG}[0]{G}
\newcommand{\floorplanV}[0]{V}
\newcommand{\floorplanE}[0]{E}

\newcommand{\shelves}[0]{S}

\newcommand{\stations}[0]{R}

\newcommand{\product}[1]{\rho_{#1}}
\newcommand{\products}[0]{\boldsymbol\rho}
\newcommand{\noproducts}[0]{n}

\newcommand{\location}[1]{\Lambda_{#1}}
\newcommand{\locations}[0]{\boldsymbol\Lambda}

\newcommand{\vertexCoords}[1]{v_{#1}}

\newcommand{\planpos}[0]{\boldsymbol\pi}
\newcommand{\pos}[2]{\pi_{#1,#2}}

\newcommand{\planholds}[0]{\boldsymbol\phi}
\newcommand{\holds}[2]{\phi_{#1,#2}}

\newcommand{\productsAt}[1]{\textsc{productsAt}(#1)}
\newcommand{\unitsAt}[2]{\textsc{unitsAt}(#1, #2)}

\newcommand{\warehouse}[0]{W}
\newcommand{\demand}[1]{w_{#1}}
\newcommand{\workload}[0]{\boldsymbol{w}}
\newcommand{\timeLimit}[0]{T}

\newcommand{\component}[1]{C_{#1}}
\newcommand{\head}[1]{\textsc{head}({#1})}
\newcommand{\tail}[1]{\textsc{tail}({#1})}
\newcommand{\inlets}[1]{\textsc{inlets}({#1})}
\newcommand{\outlets}[1]{\textsc{outlets}({#1})}

\newcommand{\tsysG}[0]{G_s}
\newcommand{\tsysV}[0]{V_s}
\newcommand{\tsysE}[0]{E_s}

\newcommand{\agentCycleSet}[0]{\Sigma}

\newcommand{\cycleTime}[0]{t_c}
\newcommand{\cyclePeriods}[0]{q_c}

\newcommand{\cyclePeriodStartTime}[0]{t_s}
  
\newcommand{\agentsInComponent}[2]{\textsc{agents}(#1, #2)}
\newcommand{\advanceT}[1]{\textsc{advanceT}(#1)}
\newcommand{\cycleIndex}[1]{\textsc{cycleI}(#1)}
\newcommand{\cycle}[1]{\textsc{cycle}(#1)}
\newcommand{\accepting}[2]{\textsc{accepting}(#1, #2)}
\newcommand{\nextV}[2]{\textsc{next}(#1, #2)}
\newcommand{\noNextV}[0]{\bot}

\newcommand{\agentFlow}[3]{f_{#1,#2,#3}}
\newcommand{\agentFlowSet}[0]{F}
\newcommand{\agentFlowIn}[2]{f^{in}_{#1,#2}}
\newcommand{\agentFlowOut}[2]{f^{out}_{#1,#2}}

\newcommand{\componentContract}[1]{\tilde{C}_{#1}}
\newcommand{\assumptions}[1]{\tilde{A}_{#1}}
\newcommand{\guarantees}[1]{\tilde{G}_{#1}}

\newcommand{\trafficSystemContract}[0]{\tilde{C}_{TS}}
\newcommand{\workloadContract}[0]{\tilde{C}_{\workload{}}}

\newcommand{\agentPaths}[1]{P_{#1}}
\newcommand{\agentPathsBijection}[0]{B_F}

\usepackage{color}

\title{Co-Design of Topology, Scheduling, and Path Planning in Automated Warehouses
\thanks{This research was supported in part by the National Science Foundation (NSF) under Awards 1846524 and 2139982, the Office of Naval Research (ONR) under Award N00014-20-1-2258, the Defense Advanced Research Projects Agency (DARPA) under Award HR00112010003, and the 2022 Okawa Research Grant. The project has also received funding from the European Union’s Horizon 2020 research and innovation program under the Marie Sk\l{}odowska-Curie grant agreement No. 894237.}}

\author{
\IEEEauthorblockN{Christopher Leet$^1$, Chanwook Oh$^1$, Michele Lora$^{1,2}$, 
Sven Koenig$^1$, Pierluigi Nuzzo$^1$}
\IEEEauthorblockA{\textit{$^1$University of Southern California, Los Angeles, California, USA}}
\IEEEauthorblockA{\textit{$^2$University of Verona, Verona, Italy}}
{\tt \footnotesize \{cjleet|chanwooo|loramich|skoenig|nuzzo\}@usc.edu}
}

\begin{document}
\maketitle

\begin{abstract}
We address the \emph{warehouse servicing problem} (WSP) in automated warehouses, which use teams of mobile agents to bring products from shelves to packing stations. Given a list of products, the WSP amounts to finding a plan for a team of agents  % \pierluigi{That works. In the first sentence we are application-oriented. In the second we are formalizing the problem. Agent is an abstraction. No big deal in any case.} 
which brings every product on the list to a station within a given timeframe. The WSP consists of four subproblems, concerning \emph{what} tasks to perform (task formulation), \emph{who} will perform them (task allocation), and \emph{when} (scheduling) and \emph{how} (path planning) to perform them. These subproblems are NP-hard individually and become more challenging in combination. The difficulty of the WSP is compounded by the scale of automated warehouses, which frequently use teams of hundreds of agents. In this paper, we present a methodology that can solve the WSP at such scales. We introduce a novel, contract-based design framework which decomposes an automated warehouse into traffic system components. By assigning each of these components a contract describing the traffic flows it can support, we can synthesize a traffic flow satisfying a given WSP instance. 
% \pierluigi{OK, I made the adjustment. I am sure we can make these sentences tighter later on. Also, we could say ``we present a methodology that leverages platform-based design with assume-guarantee contracts to solve the WSP at such scales.'' However, let's wait and see how much of this will be in the paper before adding even more verbosity.} 
Component-wise search-based path planning is then used to transform this traffic flow into a plan for discrete agents in a modular way. 
% \pierluigi{Indeed, this is a bit vague but should be fine for now. Modularity can be mentioned earlier too. See above}.  
Evaluation shows that this methodology can solve WSP instances on real automated warehouses.
%
%Automated warehouses, warehouses which use teams of mobile robots to bring products from shelves to packing stations, must solve the \emph{warehouse servicing problem} (WSP) to operate. In the WSP, we are given a list of products and asked to find a plan for a team of agents which brings every product on this list to a station within a given timeframe. The WSP consists of four subproblems, \emph{``what"} (task formulation), \emph{``who"} (task allocation), \emph{``when"} (scheduling) and \emph{``how"} (path planning), which are NP-Hard individually and more complex in combination. The difficulty of the WSP is compounded by the scale of automated warehouses, which frequently use teams of hundreds of robots. In this paper, we present the first ever methodology which can solve the WSP at such scales. We introduce a novel platform based design framework which decomposes an automated warehouse into a system of traffic system components. By assigning each of these components a contract describing the traffic flows it can support, we can synthesize a traffic flow satisfying a given WSP instance. Component-wise search-based path planning is used to transform this traffic flow into a plan for discrete agents. Evaluation shows that this methodology can solve WSP instances on real automated warehouses.
\end{abstract}

\section{Introduction}
An \emph{automated warehouse} is a warehouse which uses a team of mobile agents to move products from its shelves to its packing stations. Over the past decade, automated warehouses have become increasingly important to industrial logistics and e-commerce. Today, companies such as Amazon routinely deploy teams of hundreds of agents to manage large warehouse complexes~\cite{amazon_warehouses}. An automated warehouse must solve the \emph{warehouse servicing problem} (WSP) to operate. In the WSP, we are given a warehouse and a list of products termed a \emph{workload}, 
% (which may include duplicates) 
and asked to find a plan for a team of agents which brings every product on the list to one of the warehouse's stations within a given time limit. The WSP consists of four interdependent sub-problems:
\begin{enumerate}
    \item \emph{Task Formulation.} Which shelf and station should a product be taken and brought to? 
    \item \emph{Task Assignment.} Which tasks should an agent perform?
    \item \emph{Scheduling.} When should an agent perform its tasks?
    \item \emph{Path Planning.} What path should an agent take through the warehouse to perform each of its tasks?
\end{enumerate}
Task assignment, scheduling, and path planning are NP-Hard~\cite{ scheduling_NPHard}. Interdependence only increases the challenge of these sub-problems. As a result, existing methodologies for performing task assignment, scheduling, and path planning concurrently~\cite{MAPD1} do not scale beyond tens of agents. Automated warehouses, however, often have hundreds of agents~\cite{amazon_warehouses}. Methodologies for each solving subcase of the WSP which omit one or more of these sub-problems have also been proposed~\cite{EECBS}, but it is unclear whether any can be extended to the full WSP. The question: \emph{``is it possible to solve the WSP at scale?"} is thus open and highly relevant. 
% \pierluigi{Might be better to spell out the acronyms.} \chris{Better now?}

We answer the question in the affirmative by developing a novel traffic-system-based methodology for the WSP. In a traffic-system-based warehouse, each shelf and station is linked by a network of roads termed a \emph{traffic system}. The movement of agents through a traffic system is restricted by the rules of the traffic system. Well chosen rules prune the space of potential solutions to the WSP dramatically while preserving efficient solutions. Almost all automated warehouses today use traffic systems to plan for their agents. 

In this paper, we present the first formal framework for designing a warehouse traffic system. This framework provides a designer with a library of traffic system \emph{components} and rules for composing these components into a traffic system. There are three types of components: \emph{shelving rows}, which provide access to shelves, \emph{station queues}, which provide access to stations, and \emph{transports}, which connect other components. The traffic flows that a component can support are captured by an assume-guarantee contract termed a \emph{component contract}.

This compositional formalization of a traffic system allows for a compositional formulation of the planning problem. In this formulation, a plan is composed out of a set of \emph{agent cycles}. An agent cycle is a cycle of traffic system components containing a \emph{target shelving row} and \emph{target station queue}. The agents in an agent cycle loop through this cycle of components, carrying products from the target shelving row to the target station queue. We synthesize a plan for a WSP instance by finding a set of agent cycles which solves this instance and satisfies the constraints posed by a given traffic system. 
% can support.

Based on this compositional view of a plan, our methodology for synthesizing a plan for a WSP instance proceeds as follows.  The traffic flow required to service the instance's workload within the instance's time limit is captured by an assume-guarantee contract termed a \emph{workload contract}. A traffic flow is found which satisfies the conjunction of this workload contract with the composition of the traffic system's component contracts. This traffic flow both solves the WSP instance and can be supported by the traffic system. 

This traffic flow is then mapped to a set of agent cycles, which is converted to a plan in a modular fashion. 
% Time is discretized. 
Each timestep, a component moves each agent that it contains toward the next component in its agent cycle. Evaluations show that our methodology can solve WSP instances with hundreds of agents and thousands of tasks on real warehouse layouts in under a minute.

\section{Background}

\subsection{Prior Work}
% This paper presents the first methodology which performs the task formulation, task assignment, scheduling and path planning in automated warehouses concurrently. Prior work has, however, studied subcases of this problem.  

% The simplest subcase of this problem is the Multi-Agent Path Finding (MAPF) problem. In the MAPF problem, we given a set of agents positioned in a graph and asked to find a collision-free plan which moves each agent to its goal vertex. The aim is to minimize the makespan of the plan, the sum of the travel times of its agents. Extensive work has been done on MAPF~\cite{basic_mapf}. Simple MAPF variants include lifelong MAPF~\cite{RHCR}, where an agent must visit a sequence of goal vertices.

% Variants of MAPF involving scheduling and task assignment have been studied. Multi-Goal Sequencing and Path Finding (MGSPF), a variant of lifelong MAPF where an agent may visit its goal vertices in any order, is studied in~\cite{}. Unlabeled MAPF, where agents must be assigned goal vertices, is studied in [].  These solutions, however, have not been yet scaled beyond 100 and 35 agents respectively.  

While this paper is the first to formalize and solve the WSP problem, a related problem termed the Multi-Agent Pickup and Delivery (MAPD) problem has been studied~\cite{MAPD1}. In MAPD, we are given a set of tasks characterized by a pickup vertex $v_i$, a delivery vertex $v_j$, and a release time $t$, and asked to execute each task by moving an agent from $v_i$ to $v_j$ after time $t$. 
Studied variants include lifelong MAPD~\cite{lifelong_mapd}, where a task is not revealed until its release time, and deadline aware MAPD~\cite{deadline_mapd}, where each task has a deadline. These solvers are not directly applicable to the WSP problem, however, because they do not perform task formulation and have not been shown to scale beyond 50-75 agents.

%The Deadline Aware Multi-Agent Tour problem, a variant of MGSPF where each goal vertex has a deadline and the objective is to maximize the number of satisfied deadlines, is solved in []. This solution scales well on an open grid but it is unclear if it scales on other graph topologies.

\subsection{Assume-Guarantee Contracts}

We provide an overview of the Assume-Guarantee (A/G) contract framework by starting with the notion of a component. A component $M$ is an element of a design, which can be connected with other components to form larger systems.
An A/G contract $\tilde{C} \isdef (V, \tilde{A}, \tilde{G})$ is a triple where $V$ is a set of component variables, $\tilde{A}$ is  a set of \emph{assumptions}, that is, the set of behaviors that a component $M$ expects from the environment, and $\tilde{G}$ is the \emph{guarantees}, that is, the set of behaviors promised by the component $M$ if the assumptions hold.  A/G contracts can be combined via the \emph{composition  ($\otimes$)} or 
\emph{conjunction ($\land$)} operators. Let $\componentContract{1}$ and $\componentContract{2}$ be contracts describing the components $M_1$ and $M_2$. Taking the composition of $\componentContract{1}$  and $\componentContract{2}$ produces a contract which describes the system formed by composing the components $M_1$ and $M_2$. Taking the conjunction of $\componentContract{1}$ and $\componentContract{2}$ produces a contract which combines the requirements of the two contracts. Further details on the A/G contract framework can be found in the literature~\cite{Benveniste18}.
% A contract $C=(V,A,G)$ is in \emph{saturated form} if $G = G \cup \overline{A}$, where $\overline{A}$ is the complement of A. A contract $C=(V,A,G)$ can be \emph{saturated} by turning it into $C'=(V,A',G')$ where $A'=A$ and $G'= G \cup \overline{A}$. $C$ and $C'$ are equivalent, since they are characterized by the same sets of environments and implementations. In the remainder of the article, unless otherwise specified, we assume that all contracts are saturated, as saturated forms will be used to define the main contract operations and relations. 

% A/G contracts can be ordered according to a \emph{refinement} relationship. A contract $C=(V,A,G)$ refines $C'=(V,A',G')$, denoted as $C \preceq C'$, if and only if $A \supseteq A'$ and $G \subseteq G'$ hold. The relationship amounts to weakening the assumptions and strengthening the guarantees. Thus, if $M \models C$ and $C \preceq C'$, then $M \models C'$. We can then replace $C'$ with $C$ in the design. 

% 

% \subsection{Automated Warehouse}
% An automated warehouse uses a team of mobile agents are used to bring products from shelves to packing stations. Each agent moves at a constant speed and can carry at most one product. \multifigref{fig:warehouse_example}{a} shows the floorplan of an example warehouse containing two shelves and one station. Both shelves contain 10 units of a single product $\product{1}$. 

\section{Problem Formulation}

\begin{figure}[t]
    \centering
    \includegraphics[width=0.7\linewidth, trim={0 0 0 0},clip]{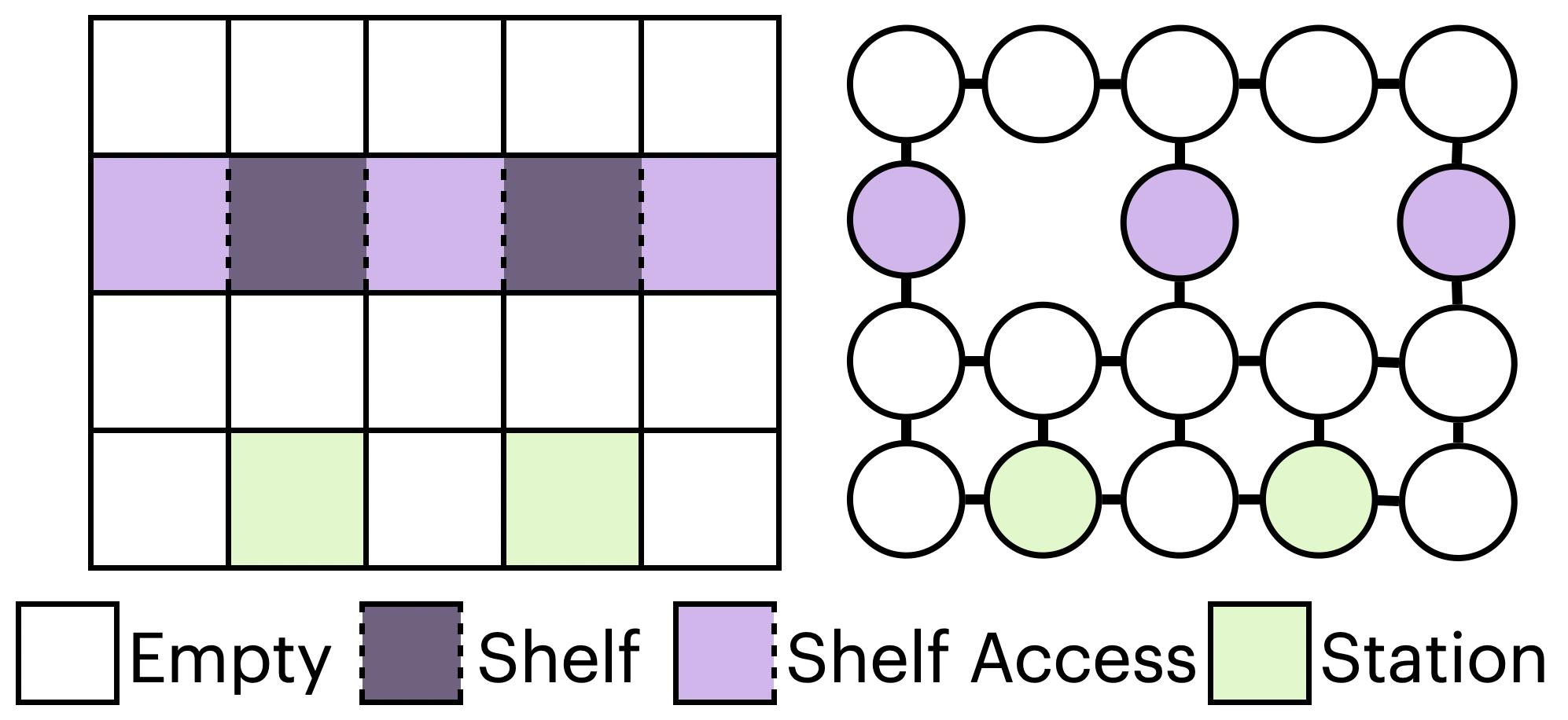}
    \caption{(left) An example warehouse and (right) its floorplan graph.}
    \vspace{-3mm}
    \label{fig:example_warehouse}
\end{figure}

An automated \emph{warehouse} $\warehouse{} \isdef (\floorplanG{}, \shelves{}, \stations{}, \products{}, \locations{})$ is represented as a 5-tuple containing the following elements:

\begin{itemize}
\item \textit{Floorplan Graph $\floorplanG{} \isdef (\floorplanV{}, \floorplanE{})$.} An undirected graph representing the warehouse's floorplan where each vertex $v_i \in \floorplanV{}$ represents a one-agent-wide cell 
% \chanwook{Is it a cell which can only hold one agent at a time? Any clearer definition of ``one-agent-wide cell''?} 
in the floorplan. There is an edge $(v_i, v_j) \in \floorplanE{}$ 
% represents a directed connection 
%between the vertices $v_i$ and $v_j$ 
if and only if (iff) an agent can move from  $v_i$ to $v_j$ without traversing another cell.
  
\item \textit{Shelf Access Vertices $\shelves{}  \subset \floorplanV{}$.} The vertices in $\floorplanV{}$ which an agent can access a shelf from.
  
\item \textit{Station Vertices $\stations{}  \subset \floorplanV{}$.} The vertices in $\floorplanV{}$ which workers can access an agent from.

\item \textit{Product Vector $\products{} \isdef \langle \product{1}, \ldots, \product{\noproducts{}} \rangle$.} The products that warehouse $W$ contains.
  
\item \textit{Location Matrix $\locations{}$.} A $|\products{}| \times |\shelves{}|$ matrix where $\locations_{k,l} \in \mathbb{N}$ is the number of units of product $\product{k}$ accessible from shelf access vertex $v_l$. 
\end{itemize}

\multifigref{fig:example_warehouse}{left} shows a warehouse with two shelves and two stations. Shelves are accessed from the east and west. \multifigref{fig:example_warehouse}{right} shows the floorplan graph $\floorplanG{} \isdef (\floorplanV{}, \floorplanE{})$ of this warehouse. If $\vertexCoords{x,y}$ is the vertex in $\floorplanV{}$ representing the cell at coordinates $(x,y)$, this warehouse has shelf access vertices $\shelves{} = \{\vertexCoords{0,2}, \vertexCoords{2,2}, \vertexCoords{4,2}\}$ and stations $\stations{} =  \{\vertexCoords{1,0}, \vertexCoords{3,0}\}$. If the shelves at $(1,2)$ and $(3,2)$ contain $10$ units of product $\product{1}$ and $\product{2}$, respectively, this warehouse has product vector $\products{} \isdef \langle \product{1}, \product{2} \rangle$ and location matrix: 
% \pierluigi{Should be 2$\times$2?} \chris{should be shelf \emph{access} vertices, corrected}
%
\begin{equation*}
\locations{} = 
\begin{bmatrix}
10 &10 &0\\
0  &10 &10
\end{bmatrix}.
\end{equation*}
  
Products are carried by a team of mobile \emph{agents} $\agents{} \isdef \langle \agent{1}, \agent{2}, \ldots \rangle$. Time in a warehouse is discretized. At each timestep $t$, an agent $\agent{i}$ has state $(\pos{i}{t}, \holds{i}{t}) \in \floorplanV{} \times \{\product{0}\} \cup \products{}$, where $\pos{i}{t}$ and $\holds{i}{t}$ are the vertex that agent $\agent{i}$ occupies and the product $\product{k} \in \products{}$ that it holds at time $t$ respectively. If agent $\agent{i}$ is not holding a product at time $t$, $\holds{i}{t} = \product{0}$. 

A \emph{$\timeLimit{}$ timestep plan} $(\planpos{}, \planholds{})$ for a team of $\noagents{}$ agents is a pair of $\noagents{} \times \timeLimit{}$ matrices such that $(\pos{i}{t}, \holds{i}{t})$ is the state of agent $\agent{i} \in \agents{}$ at time $t \in [1, \timeLimit{}]$. A $\timeLimit{}$-timestep  plan is \emph{feasible} iff:

\emph{(1)} An agent $\agent{i}$ moves by 0 or 1 vertices per timestep, that is, the vertex $\pos{i}{t+1}$ that $a_i$ occupies at step $t+1$ is the same as or adjacent to the vertex $\pos{i}{t}$ that $\agent{i}$ occupies at step $t$:
\begin{equation*}
\forall\ t \in [1,T],\ \forall\ \agent{i} \in \agents{},\ \pos{i}{t+1} \in \{ \pos{i}{t} \} \cup Adj(\pos{i}{t}).
\end{equation*}

\emph{(2)} Two agents do not collide, that is, two agents $\agent{i}, \agent{j} \in \agents{}$ do not occupy the same vertex or traverse the same edge in opposite directions at the same timestep:
\begin{align*}
\forall\ t \in [&1, T],\ \forall\ \agent{i},\agent{j} \in \agents{},\\ 
&\pos{i}{t} \neq \pos{j}{t} \wedge \neg(\pos{i}{t+1} = \pos{j}{t} \wedge \pos{i}{t} = \pos{j}{t+1}).
\end{align*}

\emph{(3)} An agent can only pick up a product $\product{k}$ at a shelf access vertex containing $\product{k}$ and put down a product at a station. Let $\productsAt{v}$ be the set of products accessible at vertex $v$ (if $v \notin S$, $\productsAt{v} = \emptyset$); then, %
\begin{align*}
&\forall\ t \in [1, T],\ \forall\ \agent{i} \in \agents{},\\%
&\holds{i}{t+1} \in %
\begin{cases}
\{\product{0}\} \cup \productsAt{\pos{i}{t}} %
 &\holds{i}{t} = \product{0}\\ %
\{\product{0}, \holds{i}{t}\}                         
 &\pos{i}{t} \in \stations{}\\ %
\{\holds{i}{t}\} %                         
 &\text{otherwise}.
\end{cases} 
\end{align*}
% \pierluigi{$\holds{i}{t-1}$ or $\holds{i}{t}$ in the first case?}\chris{fixed}

A \emph{workload} 
% $\workload{} \isdef \langle \demand{1}, \ldots, \demand{\noproducts{}} \rangle$ 
is a vector $\workload{} \isdef \langle \demand{1}, \ldots, \demand{\noproducts{}} \rangle$ where $\demand{k}$ is the number of units of product $\product{k}$ which must be transferred to a station. We say that a  $\timeLimit{}$ timestep plan \emph{services} workload $\workload{}$ iff it is feasible and it transfers $\demand{k}$ units of each product $\product{k} \in \products{}$ to the warehouse's stations $\stations{}$. 

\begin{problem}[Warehouse Servicing Problem]
Given a warehouse $\warehouse{}$, a workload $\workload{}$, and a timestep limit $T$, find a $\timeLimit{}$ timestep plan containing an arbitrary number of agents which services workload $\workload{}$. 
\end{problem}

\section{Co-Design Methodology}

\begin{figure}[t]
    \centering
    \includegraphics[width=0.8\linewidth, trim={0 0 0 0},clip]{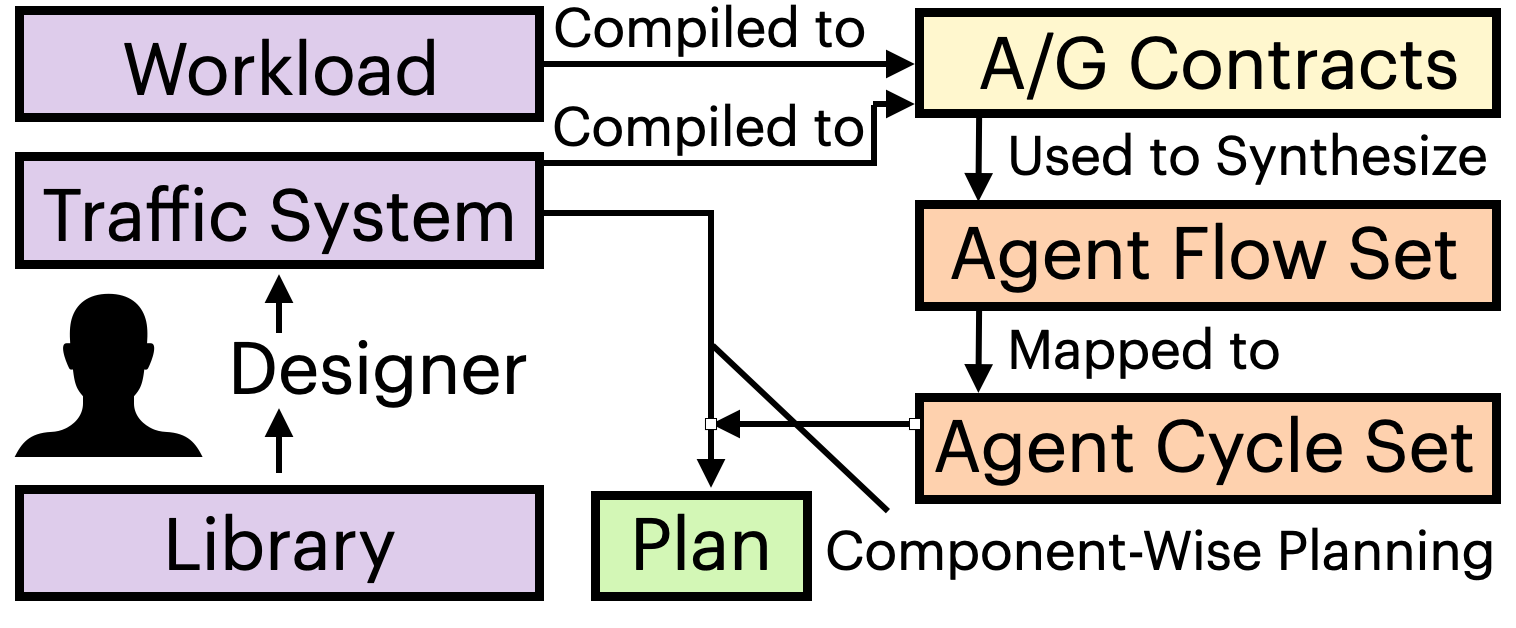}
    \vspace{-2mm}
    \caption{The high-level workflow of the methodology.}
    \vspace{-5mm}
    \label{fig:workflow}
\end{figure}

The structure of the methodology is shown in ~\figref{fig:workflow}, providing a framework for designing a traffic system 
% to a designer
(Subsection~\ref{subsec:traffic_system_design}). The framework offers rules for grouping the vertices in a warehouse floorplan graph $\floorplanG{}$ into traffic system components. The high-level movement of agents through a traffic system is represented by an agent cycle set (Subsection~\ref{subsec:agent_cycle_set}). An agent cycle set is converted into a plan in a modular fashion (Subsection~\ref{subsec:realizing_an_agent_cycle_set}), such that, 
% Time is discretized. 
at each timestep, each component moves each agent it contains toward the next component in that agent's agent cycle set.

 The methodology finds an agent cycle set for a traffic system which services a workload $\workload{}$ within $\timeLimit{}$ timesteps as follows. The properties of the flow of agents entering and leaving a component
 % assumes that  it has certain properties and guarantees that the flow of agents  it has certain properties. These assumptions and guarantees 
 are compiled into an A/G contract. The properties that a traffic system's set of agent flows must have to service workload $\workload{}$ within $\timeLimit{}$ timesteps are also compiled into an A/G contract. An agent flow set satisfying the conjunction of these contracts is found (Subsection~\ref{subsec:synthesizing_an_agent_cycle_set}) and mapped to an agent cycle set (Subsection~\ref{subsec:agent_flow_set_to_agent_cycle_set}). By construction, the resulting agent cycle set is supported by the traffic system.

%\pierluigi{Would be good to have an introductory paragraph with an intuitive overview/flow of the entire methodology? Maybe also modify Fig. 2 to show the entire methodology? Also, it would be good to connect to the title, showing where the topology design, scheduling, and planning happen. I guess this can be done also with the help of the intro.}

\subsection{Traffic System Design Framework}
\label{subsec:traffic_system_design}

An operator can construct a \emph{traffic system} for a warehouse by dividing the vertices in its floorplan graph into disjoint simple paths called traffic system \emph{components}. A component $\component{i}$ behaves similarly to a one way road. Agents enter a component $\component{i}$ at its head $\head{\component{i}}$ and exit it from its tail $\tail{\component{i}}$.  A component may not contain both shelf access vertices and station access 
vertices. A component is termed a:
\begin{enumerate}
    \item \emph{shelving row} if it contains shelf access vertices;
    \item \emph{station queue} if it contains station vertices;
    \item \emph{transport} if it contains neither.
\end{enumerate}
% \pierluigi{Have we defined station access vertices?} \chris{corrected}
Every station and shelf access vertex must be contained by a component. Other vertices, however, need not be. Vertices which are not part of any component are termed \emph{unused vertices} since they will not be traversed by any agent.

A component $\component{i}$ has 1 or 2 \emph{inlet components} $\inlets{\component{i}}$ and 1 or 2 \emph{outlet components} $\outlets{\component{i}}$. Agents must enter $\component{i}$ from one of its inlets and exit $\component{i}$ from one of its outlets.  There must be an edge in the floorplan graph between the head of a component and the tail of each of its inlets and the tail of a component and the head of each of its outlets.

The connections between the components of a traffic system are represented by a directed graph termed a \emph{traffic system} graph $\tsysG \isdef (\tsysV, \tsysE)$. Each vertex in $\tsysV$ is a component of the traffic system. There is an arc $(\component{i}, \component{j})$ in $\tsysV$ iff component $\component{i}$ is one of component $\component{j}$'s inlets. A traffic system graph must be strongly connected, that is, there must be a path between any two components in a traffic system graph. Since every shelf access and station access vertex is in a component, it follows that there is a way for an agent to move from any shelf access vertex to any station access vertex and \emph{vice versa}.

\subsection{Agent Cycle Set}
\label{subsec:agent_cycle_set}

We synthesize a plan for a traffic system $\tsysG$ which satisfies a workload $\workload{}$ within $\timeLimit{}$ timesteps from an \emph{agent cycle set} $\agentCycleSet{}$. An \emph{agent cycle} is a set of $b$ agents associated with a cycle of $b$ components in the traffic system graph $\tsysG{}$. These components must include a \emph{target shelving row} and a \emph{target station queue}. The agents in an agent cycle transfer products from the \emph{target shelving row} to the \emph{target station queue}.

An agent cycle set has a \emph{cycle time} $\cycleTime{}$. At timestep $t=1$, each agent in an agent cycle is positioned on a unique component. Every $\cycleTime{}$ timesteps, each agent in an agent cycle advances one component. Thus, an agent cycle delivers one product from its target shelving row to its target station queue every $\cycleTime{}$ timesteps. In Subsection~\ref{subsec:realizing_an_agent_cycle_set} we show how we find a plan that realizes the agent movement specified by an agent cycle set. (Due to space constraints, we do not specify the exact timesteps at which an agent cycle picks up and drops off products at its target shelving row and target station queue.)  In Subsection~\ref{subsec:synthesizing_an_agent_cycle_set}, we show how we find an agent cycle set that services a workload $\workload{}$ within $\timeLimit{}$ timesteps.

\subsection{Realizing an Agent Cycle Set}
\label{subsec:realizing_an_agent_cycle_set}

Let the time interval $[1, \timeLimit{}]$ be divided into the $\lfloor \timeLimit{}/\cycleTime{} \rfloor$ \emph{cycle periods} $[1, \cycleTime{}]$, $[\cycleTime{}+1, 2\cycleTime{}]$, etc. Let $|\component{i}|$ be the number of vertices in a component $\component{i}$ and $m$ be the length of the longest component in a traffic system, i.e., 
\begin{equation*}
m \isdef \max(|\component{i}| : \component{i} \in \tsysV{}).
\end{equation*}
The realization algorithm has the following property:
\begin{property}
\label{subsec:realization_condition}
Our realization algorithm can realize an agent cycle set $\Sigma$ if it has cycle time $2m$ and there is no component $\component{i}$ contained by more than $\lfloor |\component{i}|/2 \rfloor$ agent cycles.
\end{property}

The realization algorithm moves an agent at least once every other timestep until the agent advances to the next component in its agent cycle. It therefore takes at most $2m$ timesteps for the realization algorithm to advance any agent in any agent cycle by one component. The realization algorithm assumes, however, that a component $\component{i}$ can send an agent to its outlets at least once every other timestep. This condition is violated if a component's outlets fill up with agents, preventing them from accepting additional agents. To prevent a component from filling up, the realization algorithm prevents agents from advancing multiple components in a single cycle period. Since each component $\component{i}$ is in at most $\lfloor |\component{i}|/2 \rfloor$ agent cycles, no more than $2 \cdot \lfloor |\component{i}|/2 \rfloor$ unique agents will occupy any component $\component{i}$ every cycle period. Since $2 \cdot \lfloor |\component{i}|/2 \rfloor \leq |\component{i}|$, no component will ever fill up.

\para{Realization Algorithm Initialization} At timestep $t=1$, we place an agent associated with an agent cycle on an arbitrary vertex of each component that the agent cycle passes through.

\para{Realization Algorithm Timestep} We realize the location of the agents in 
a traffic system at time $t+1$ from their locations at time $t$ by calling the function $\textsc{ComponentTimestep}(\component{i}, t, \pos{1}{t}, \pos{2}{t}, \ldots)$, Algorithm~\ref{alg:component_timestep}, on every component $\component{i}$ in the traffic system.

\begin{algorithm}[t]
\small
\caption{\small\textsc{ComponentTimestep}($\component{i}$, $t+1$, $\pos{1}{t}$, $\pos{2}{t}$, $\ldots$)}
\label{alg:component_timestep}
\begin{algorithmic}[1]

\State $t_s \gets \textsc{CyclePeriodStartT}(t)$
\label{ln:cyclePeriodStartT}

\State $a_j \gets \head{\agentsInComponent{\component{i}}{t}}$ 
\label{ln:get_head_agent}

\If{$\pos{j}{t} = \head{\component{i}} \wedge \advanceT{\agent{j}} < \cyclePeriodStartTime{}$}
\label{ln:advance_check}

  \State $k \gets \cycleIndex{\agent{j}} + 1 \mod |\cycle{\agent{j}}|$
  \label{ln:compute_cycle_index}  

  \If{$\accepting{\component{i}}{\cycle{\agent{j}}[k]}$}

    \State $\pos{j}{t+1} \gets \tail{\cycle{\agent{j}}[k]}$
    \label{ln:advance_agent_st}
    \State $\cycleIndex{\agent{j}} \gets k$
    \State $\advanceT{\agent{j}} \gets t+1$
    \label{ln:advance_agent_en}
  \EndIf  
\EndIf

\For{$\agent{j} \in \agentsInComponent{\component{i}}{t}$}
\label{ln:internal_mv_st}

  \State $v \gets  \nextV{\component{i}}{\pos{j}{1}}$

  \If{$v \neq \bot \wedge \neg \exists\ \agent{k} \in \agentsInComponent{\component{i}}{t} : \pos{k}{t} = v$}
    \State $\pos{j}{t+1} \gets v$
  \Else
    \State $\pos{j}{t+1} \gets \pos{j}{t}$
  \EndIf
\EndFor
\label{ln:internal_mv_en}
\end{algorithmic}
\end{algorithm}

Let $\cyclePeriodStartTime{}$ be the time at which the current cycle period starts (Line~\ref{ln:cyclePeriodStartT}),  $\agentsInComponent{\component{i}}{t}$ be a list of the agents in $\component{i}$ at time $t$ ordered by their distance from $\head{\component{i}}$, and $\agent{j}$ be the agent at the head of this list (Line~\ref{ln:get_head_agent}). If $\agent{j}$ is at the head of $\component{i}$ and $\advanceT{\agent{j}}$, the timestep when $\agent{j}$ advanced to $\component{i}$, was before $\cyclePeriodStartTime{}$, we check if $\agent{j}$ can advance to the next component in its agent cycle $\cycle{\agent{j}}$ (Line~\ref{ln:advance_check}). We compute the index $k$ of the next component in $\cycle{\agent{j}}$ from the index $\cycleIndex{\agent{j}}$ of the current component in $\cycle{\agent{j}}$ (Line~\ref{ln:compute_cycle_index}). If the component $\cycle{\agent{j}}[k]$ is accepting agents from $\component{i}$, we move agent $\agent{j}$ to this component's tail (Lines~\ref{ln:advance_agent_st}-\ref{ln:advance_agent_en}). Next, we move agents within $\component{i}$. Let $\nextV{\component{i}}{u}$ be the vertex in $\component{i}$ following vertex $u$, if one exists, and $\noNextV$ otherwise. We move each agent $\agent{j} \in \agentsInComponent{\component{i}}{t}$ to vertex $\nextV{\component{i}}{\pos{j}{1}}$ if it exists and is not occupied by another agent (Lines~\ref{ln:internal_mv_st}-\ref{ln:internal_mv_en}).

\subsection{Synthesizing an Agent Flow Set}
\label{synthesizing_an_agent_flow_set}

\begin{figure}[t]
    \centering
    \includegraphics[width=0.9\linewidth, trim={0 0 0 0},clip]{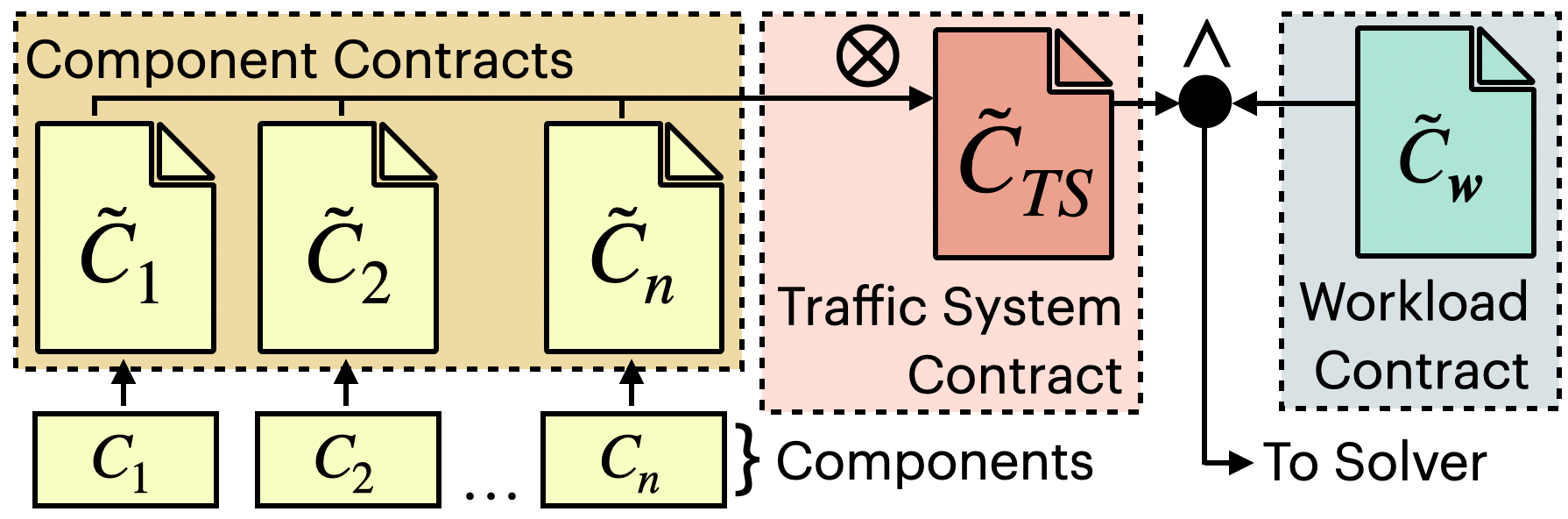}
    \caption{Synthesizing agent flows using contracts.}
    \vspace{-5mm}
    \label{fig:flow_synthesis}
\end{figure}

\label{subsec:synthesizing_an_agent_cycle_set}
Let an \emph{agent flow} $\agentFlow{i}{j}{k} \in \{0\} \cup \mathbb{N}$ be the number of agents  that move from component $\component{i}$ to component $\component{j}$ carrying product $\product{k}$ every $\cycleTime{}$ timestep period in a plan. Let an \emph{agent flow set} $\agentFlowSet{}$ be the set containing all of a plan's agent flows:
\begin{equation*}
    \agentFlowSet{} \isdef \{\agentFlow{i}{j}{k} : (\component{i}, \component{j}) \in \tsysE{} \wedge \product{k} \in \products{}\}.
\end{equation*}
An agent flow set completely describes how agents move between the components in a traffic system each cycle period. We synthesize an agent flow set which services workload $\workload{}$ within $\timeLimit{}$ timesteps as follows.

% \pierluigi{Maybe make this a subsubsection if the contracts are paragraphs?}  \pierluigi{There may be a conceptual jump here. How does it happen? Do you want to say that you instantiate/formulate contracts specifying the workload and traffic system requirements/properties as well as the agent flow constraints? Then a solution satisfying all these constraints is found in terms of agent flow sets, and these are eventually mapped to agent cycle sets that are guaranteed to be implementable by Sec. IV-C? I am trying to put this section in perspective based on the overall flow.}

A component $\component{i}$ assumes that the agent flows entering it have certain properties and guarantees that the agent flows leaving it have certain properties. These assumptions and guarantees are compiled into an A/G contract $\componentContract{i} \isdef (\assumptions{i}, \guarantees{i})$,  termed a \emph{component contract} (\figref{fig:flow_synthesis}, yellow). Component contracts are composed into a \emph{traffic system contract} $\trafficSystemContract{}$ (\figref{fig:flow_synthesis}, red) describing the agent flow sets that the traffic system allows, i.e., 
% \pierluigi{indeed, Fig. 2 should be fixed to reflect this. Also, can it be twisted a bit to pictorially represent the cycles etc. so we give a view of the whole methodology?}
%
\begin{equation*} 
\trafficSystemContract{} \isdef \bigotimes_{\component{i} \in \tsysV{}} \componentContract{i}.
\end{equation*}

The properties that a traffic system's agent flow set must have to service workload $\workload{}$ within $\timeLimit{}$ timesteps are compiled into an A/G contract termed a \emph{workload contract} $\workloadContract{}$ (\figref{fig:flow_synthesis}, blue).  We attempt to synthesize an agent flow set which satisfies the conjunction of the traffic system contract and workload contract $\workloadContract{}$ (\figref{fig:flow_synthesis}). If no such agent flow set exists, a plan which services workload $\workload{}$ within $\timeLimit{}$ timesteps cannot be synthesized using our methodology. 

\para{Component Contract Assumptions} As discussed in Subsection~\ref{subsec:realizing_an_agent_cycle_set}, at most $\lfloor |\component{i}|/2 \rfloor$ agents may enter any component $\component{i} \in \tsysV{}$ in any cycle period, i.e., 
\begin{equation*}
\forall\ \component{i} \in \tsysV{},\ \sum_{\mathclap{\component{j} \in \inlets{\component{i}}\qquad}}\quad  \sum_{\product{k} \in \products{}} \agentFlow{j}{i}{k} \leq \left \lfloor \frac{|\component{i}|}{2} \right \rfloor. 
\end{equation*}
Since a flow $\agentFlow{i}{j}{k}$ is defined as a non-negative integer, the minimum flow entering any component $\component{i} \in \tsysV{}$ is 0. 

\para{Component Contract Guarantees} Let $\agentFlowOut{i}{k} \in \{0\} \cup \mathbb{N}$ be the number of units of product $\product{k}$ transferred from an agent to a station in component $\component{i}$ each cycle period.  If component $\component{i}$ does not contain a station, $\agentFlowOut{i}{k} = 0$. If component $\component{i}$ contains stations, $\agentFlowOut{i}{k}$ is between 0 and the number of agents entering component $\component{i}$ carrying $\product{k}$ each cycle period, that is, 
\begin{equation*}
\agentFlowOut{i}{k} \in 
\begin{cases}
\{0\} &|\component{i} \cap \stations{}| = 0\\
[0, \sum_{\component{j} \in \inlets{\component{i}}} \agentFlow{j}{i}{k}] &\text{otherwise}
\end{cases}.
\end{equation*}

Let $\agentFlowIn{i}{k} \in \{0\} \cup \mathbb{N}$ be the number of units of product $\product{k}$ transferred from a shelf in component $\component{i}$ each cycle period. If component $\component{i}$ does not contain shelves, $\agentFlowIn{i}{k} = 0$. If component $\component{i}$ contains shelves, $\agentFlowIn{i}{k}$ is limited by the number of units of product $\product{k}$ that component $\component{i}$ contains. Let $\unitsAt{\component{i}}{\product{k}}$ be the total number of units of product $\product{k}$ available at $\component{i}$:
\begin{equation*}
\unitsAt{\component{i}}{\product{k}} \isdef \sum_{\mathclap{v_j \in \component{i} \cap \shelves{}}} \location{k,j}.
\end{equation*}

Let $\cyclePeriods{} \isdef \lfloor \cycleTime{} / \timeLimit{} \rfloor$ be the number of cycle periods executable in $\timeLimit{}$ timesteps. In $\cyclePeriods{}$ cycle periods, a component can transfer at most $\unitsAt{\component{i}}{\product{k}} / \cyclePeriods{}$ units of product $\product{k}$ each cycle period. It follows that:
\begin{equation*}
\agentFlowIn{i}{k} \in 
\begin{cases}
\{0\} &|\component{i} \cap \shelves{}| = 0\\
[0, \unitsAt{\component{i}}{\product{k}} / q] &\text{otherwise}
\end{cases}.
\end{equation*}

A product can only be transferred to an unburdened agent. Thus, the total number of products transferred to agents in $\component{i}$ each cycle period is limited by the total number of unburdened agents entering $\component{i}$ each cycle period.
\begin{equation*}
\sum_{\mathclap{\product{k} \in \products{}}} \agentFlowIn{i}{k} \leq \sum_{\mathclap{C_j \in \inlets{\component{i}}}} \agentFlow{j}{i}{0}.
\end{equation*}

Agents cannot appear or disappear. Thus, the total flow of agents carrying product $\product{k}$ out of component $\component{i}$ is:
\begin{enumerate}
\item the total flow of agents carrying product $\product{k}$ into $\component{i}$,
\item \emph{plus} the total number of unburdened agents given a unit of product $\product{k}$ from a shelf in $\component{i}$ each cycle period, 
\item \emph{minus} the total number of agents which transfer a unit of product $\product{k}$ to a station in $\component{i}$ each cycle period. Overall, 
\end{enumerate}
\begin{align*}
\forall\ \component{i}, \product{k} \in \tsysV{} \times \products{},\  \sum_{\mathclap{C_j \in \outlets{\component{i}}} \quad} \agentFlow{i}{j}{k} = \sum_{\qquad \mathclap{C_j \in \inlets{\component{i}}}} \agentFlow{j}{i}{k} + \agentFlowIn{i}{k} - \agentFlowOut{i}{k}.
\end{align*}
An analogous expression can be written relating the total flow of unburdened agents leaving and entering a component $\component{i}$:
\begin{align*}
\forall\ \component{i} \in \tsysV{},\  \sum_{\mathclap{C_j \in \outlets{\component{i}}} \quad} \agentFlow{i}{j}{0} = \sum_{\qquad \mathclap{C_j \in \inlets{\component{i}}}} \agentFlow{j}{i}{0}  + \sum_{\mathclap{\product{k} \in \products{}}} \agentFlowIn{i}{k} - \sum_{\mathclap{\product{k} \in \products{}}} \agentFlowOut{i}{k}.
\end{align*}

\para{Workload Contract} A workload contract $\workloadContract{}$ makes no assumptions. It guarantees that the total number of units of product $\product{k}$ transferred to the warehouse's stations each cycle period is greater than $\demand{k}/\cyclePeriods{}$, where $\demand{k}$ is the demand for $\product{k}$ in workload $\workload{}$ and $\cyclePeriods{}$ is the number of cycle periods $\cyclePeriods{}$ executable in $\timeLimit{}$ timesteps, i.e., 
\begin{equation*}
\forall\ \product{k} \in \products{},\ \sum_{\mathclap{\component{i} \in \tsysV{}}} \agentFlowOut{i}{k} \geq \frac{\demand{k}}{\cyclePeriods{}}.
\end{equation*}

The above contracts (and associated constraints) are used to generate a formula in propositional logic augmented with arithmetic constraints over the reals, which is solved using a satisfiability modulo theory (SMT) solver to produce the flow rate through every component for every product.

\subsection{Mapping the Agent Flow Set to an Agent Cycle Set} 
\label{subsec:agent_flow_set_to_agent_cycle_set}

By construction, an agent flow set $\agentFlowSet{}$ has the properties:

\begin{property}
\label{property:burdened_agent_paths}
There is a set of paths $\agentPaths{k}$ on $\tsysG{}$ for each product $\product{k}$ such that:
\begin{enumerate}
    \item There are exactly $\agentFlowIn{i}{k}$ paths in $\agentPaths{k}$ beginning at each component $\component{i} \in \tsysV{}$.
    \item There are exactly $\agentFlowOut{i}{k}$ paths in $\agentPaths{k}$ ending at each component $\component{i} \in \tsysV{}$.
    \item There are exactly $\agentFlow{i}{j}{k}$ paths that contain the edge $(\component{i}, \component{j})$ for each edge $(\component{i}, \component{j}) \in \tsysE{}$.
\end{enumerate}
\end{property}

\begin{property}
\label{property:unburdened_agent_paths}
There is a set of paths $\agentPaths{0}$ on $\tsysG{}$ such that:
\begin{enumerate}
\item There are exactly $\sum_{\product{k} \in \products{}} \agentFlowOut{i}{k}$ paths in $\agentPaths{0}$ beginning at each component $\component{i} \in \tsysV{}$.

\item There are exactly $\sum_{\product{k} \in \products{}} \agentFlowIn{i}{k}$ paths in $\agentPaths{0}$ ending at each component $\component{i} \in \tsysV{}$.

\item There are exactly $\agentFlow{i}{j}{0}$ paths that contain the edge $(\component{i}, \component{j})$  for each edge $(\component{i}, \component{j}) \in \tsysE{}$.
\end{enumerate}
\end{property}

Properties~\ref{property:burdened_agent_paths} and \ref{property:unburdened_agent_paths} imply that there is a bijection $\agentPathsBijection : \agentPaths{0} \rightarrow \bigcup_{\product{k} \in \products{}} \agentPaths{k}$ for any agent flow set $\agentFlowSet{}{}{}$ such that if path $p \in \agentPaths{0}$ is mapped to path $p' \in \bigcup_{\product{k} \in \products{}} \agentPaths{k}$, then the head of path $p$ is the tail of path $p'$ and \textit{vice versa}:
\begin{align*}
 \agentPathsBijection(p) = p' \Rightarrow \head{p} = \tail{p'} \wedge \head{p'} = \tail{p}.
\end{align*}

An agent cycle set $\agentCycleSet{}$ is then formed from an agent flow set $\agentFlowSet{}$ by turning each pair of paths in the bijection $\agentPathsBijection{}$ into a cycle. 

\section{Evaluations}

\begin{figure}[t]
    \centering
    \includegraphics[width=\linewidth, trim={0 0 0 0},clip]{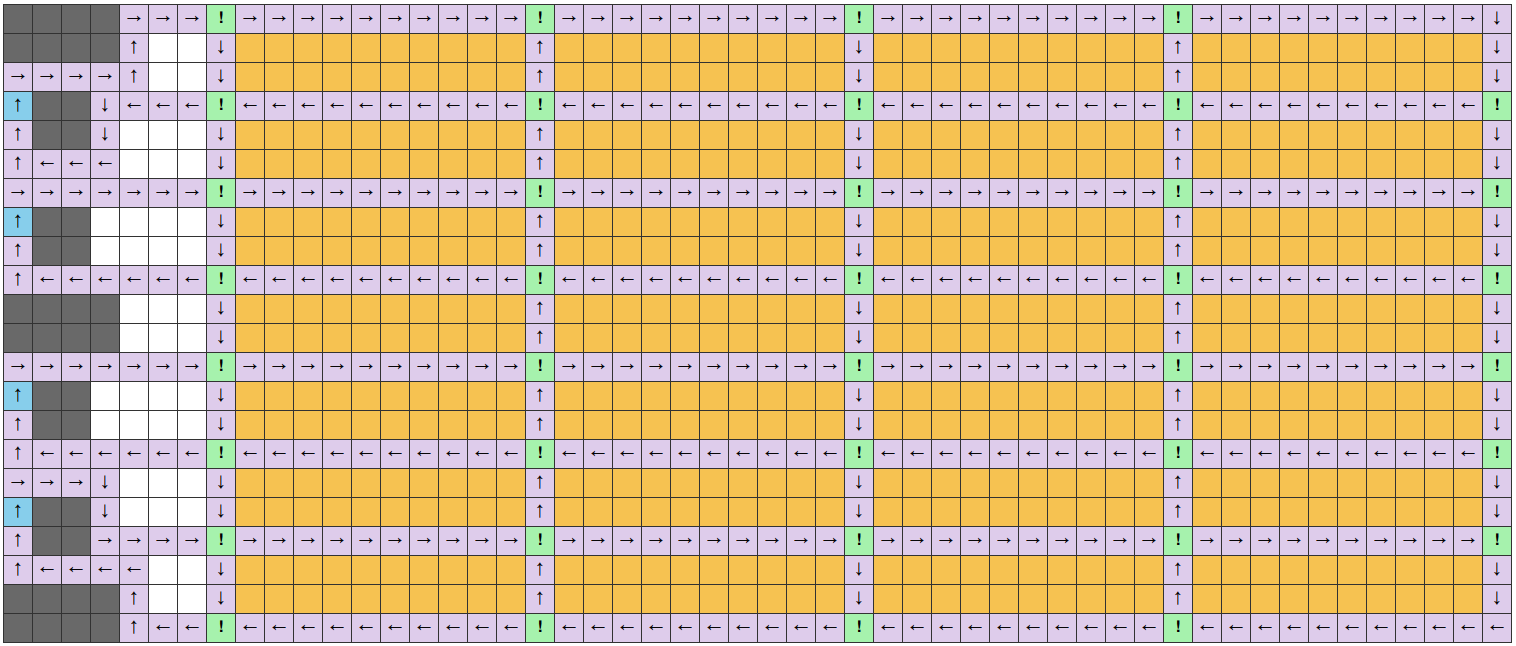}
    \caption{Fulfillment Center Map.}
    \label{fig:fulfillment_center}
\end{figure}

The proposed methodology is implemented as an automatic toolchain. The component and workload contracts are compiled and composed using the CHASE framework~\cite{CHASE}. An agent flow set satisfying these contracts is then found using Z3~\cite{Z3}. All other toolchain components  are implemented in Python 3.9. The methodology is evaluated on two real industrial scenarios taken from the literature: a Kiva (now Amazon Robotics) fulfillment center~\cite{Wurman} and a package sorting center~\cite{Wan}.

\para{Fulfillment Center} A fulfillment center ships products to individual consumers.  A fulfillment center map is characterized by blocks of shelves in its center and stations on its perimeter. The proposed approach is evaluated on two fulfillment center maps, a real map borrowed from~\cite{Wurman} with 1071 cells, 560 shelves, 4 stations, and 55 unique products and a synthetic map based on~\cite{Wurman} with 793 cells, 240 shelves, 1 station, and 120 products. The real fulfillment center map is depicted in \figref{fig:fulfillment_center}. Shelves are depicted as yellow cells, stations as blue shelves, obstacles as grey cells and empty space as white cells. The tail of each component is depicted as a green cell with an exclamation mark. Every other vertex in a component is depicted as a purple cell with an arrow pointing to the next vertex in the component.

\para{Sorting Center} The methodology is also evaluated on a variant of the WSP which takes place in a sorting center~\cite{Wan}. A sorting center sorts packages by destination. A sorting center map is characterized by uniformly placed chutes in its center and bins of unsorted packages on its perimeter. Each chute leads to a shipping counter bound for a unique destination. An agent sorts a package by ferrying it from a bin to the chute associated with its destination. Typically, a bin is modeled as having an unlimited number of packages. The goal is to fill each shipping container before it is scheduled to depart.

This problem is modeled as an WSP instance as follows. Let the $i$th chute be modeled as a shelf containing an arbitrary amount of the product $\product{i}$. Let each bin of unsorted products be modeled as a station. Let $n_i$ be the number of packages that must be brought to the $i$th chute. An instance of the WSP is generated where the demand for product $\product{i}$ is $n_i$. Solving this WSP instance produces an agent cycle set where $n_i$ units of product $\rho_i$ are brought from the $i$th chute to the bins of unsorted products. Swapping the locations where agents pick up and drop off products generates the desired solution.

\begin{figure}[t]
    \centering
    \includegraphics[width=\linewidth, trim={0 0 0 0},clip]{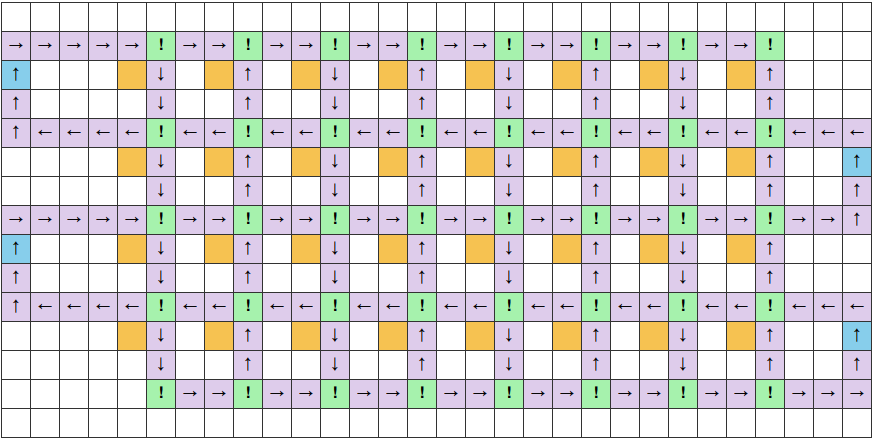}
    \caption{Sorting Center Map.}
    \label{fig:sorting_center}
\end{figure}

The sorting center evaluation is conducted on a map based on~\cite{Wan}. This map contains 406 cells, 32 chutes and, 4 bins. It is depicted in \figref{fig:sorting_center}.

\para{Experimental Hardware} Each evaluation was performed on an 2.6 GHz Intel(R) Core i7-10705H CPU with 32 GB of RAM in a Ubuntu 20.04 VM run on Windows 11.

\begin{table}[t]
    \centering
    \caption{Benchmarking the methodology on 9 WSP instances.}
    \begin{tabular}{l|l|l|l}
    \textbf{Map}  
    &\textbf{Unique Products}
    &\textbf{Units Moved}
    &\textbf{Runtime (s)}\\
    \hline
    \hline

    &36
    &160
    &8.054\\
    Sorting Center
    &36
    &320
    &8.343\\
    &36
    &480
    &14.437\\
    \hline
    &55
    &550
    &6.939\\
    Fulfillment 1
    &55
    &825
    &7.001\\
    &55
    &1100
    &8.014\\
    \hline
    &120
    &1200
    &65.880\\

    Fulfillment 2
    &120
    &1320
    &65.886\\

    &120
    &1440
    &67.825\\
    \hline
    \end{tabular}
    \label{tab:results}
    \vspace{-5mm}
\end{table}

%550 units of products:  6.939
%825 units of products:  7.001
%1100 units of products: 8.014
%Sorting center, multiple stations, 3600 units of time
%160 units of products: 8.054 seconds
%320 units of products: 8.343 seconds
%480 units of products: 14.437 seconds
%Warehouse 20x3, single station, 3600 units of time
%1200 units of products: 65.880
%1320 units of products: 65.886
%1440 units of products: 67.825

\para{Results} The three WSP instances were generated on each map.
%Each WSP instance had a time limit of 3600 seconds. 
For each WSP instance, Table~\ref{tab:results} lists:
\begin{enumerate}
    \item the number of unique products placed in the warehouse.
    \item the total units of product moved to a station.
    \item the time required to generate an agent flow set (the time required to convert an agent flow set into a plan is small).
\end{enumerate}
The length $\timeLimit$ of a plan for each WSP instance was limited to 3,600 timesteps. Solver runtime was limited to 1 hour.

The proposed methodology was able to solve an WSP instance where more than 1400 products had to be moved to a station in just over a minute. We benchmarked the methodology on this instance against Iterated EECBS~\cite{EECBS}, a state-of-the-art search-based lifelong path planner as follows. Iterated EECBS was given the start position of each agent in our solution. It was asked to find a plan where each agent visited the same sequence of shelves and stations as it did in our solution. Iterated EECBS failed to terminate after an hour.

The runtime of traditional multi-agent path planners is exponential to the size of their team of agents, the number of locations that an agent has to visit, and the average distance between the locations that an agent has to visit. Our proposed methodology, by contrast, is exponential in the number of components in the traffic system (finding a set of agent flows is reducible to the Integer Linear Programming problem). As a result, our methodology outscales state-of-the-art path planners. Additionally, our methodology is relatively insensitive to the number of products in a WSP instance. On both the sorting center and fulfillment center, doubling the units of product in the workload increased runtime by less than 10\%.

\section{Conclusion}
In this paper we introduce the first methodology for solving the warehouse servicing problem. This methodology provides a designer with a formal framework for designing a warehouse traffic system. The traffic flows that each component can support are captured using contracts. Combining these contracts with a contract capturing the traffic flow that a workload for a warehouse requires allows us to synthesize a traffic flow satisfying a given WSP instance. A simple deterministic algorithm is used to convert this traffic flow into a plan for discrete agents. The proposed methodology is evaluated on maps taken from real automated warehouses and shown to be able to solve WSP instances on these maps involving more than a thousand products in just over a minute. In future work, we hope to find a bounded-suboptimal solution to the WSP. In particular, we intend to look at ways to iteratively refine our feasible solution into an optimal solution. 

\bibliographystyle{IEEEtran}
\bibliography{references}

\end{document}